\documentclass[english,prl,reprint]{revtex4}
\usepackage[T1]{fontenc}
\usepackage[latin9]{inputenc}
\usepackage{graphicx}
\usepackage{esint}

\makeatletter
\@ifundefined{textcolor}{}
{%
 \definecolor{BLACK}{gray}{0}
 \definecolor{WHITE}{gray}{1}
 \definecolor{RED}{rgb}{1,0,0}
 \definecolor{GREEN}{rgb}{0,1,0}
 \definecolor{BLUE}{rgb}{0,0,1}
 \definecolor{CYAN}{cmyk}{1,0,0,0}
 \definecolor{MAGENTA}{cmyk}{0,1,0,0}
 \definecolor{YELLOW}{cmyk}{0,0,1,0}
 }

\makeatother

\usepackage{babel}
\begin{document}

\title{The Sign Problem, $\mathcal{PT}$ Symmetry and Abelian Lattice Duality}

\author{Peter N. Meisinger and Michael C. Ogilvie}

\affiliation{Dept. of Physics, Washington University, St. Louis, MO, USA 6313}

\email{pnm@physics.wustl.edu, mco@physics.wust.edu}

\date{06/06/13}
\begin{abstract}
Lattice field theories with complex actions are not easily studied
using conventional analytic or simulation methods. However, a large
class of these models are invariant under $\mathcal{CT}$, where $\mathcal{C}$
is charge conjugation and $\mathcal{T}$ is time reversal, including
models with non-zero chemical potential. For Abelian models in this
class, lattice duality maps models with complex actions into dual
models with real actions. For extended regions of parameter space,
calculable for each model, duality resolves the sign problem for both
analytic methods and computer simulations. Explicit duality relations
are given for models for spin and gauge models based on $Z(N)$ and
$U(1)$ symmetry groups. The dual forms are generalizations of the
$Z(N)$ chiral clock model and the lattice Frenkel-Kontorova model,
respectively. From these equivalences, rich sets of spatially-modulated
phases are found in the strong-coupling region of the original models.
\end{abstract}


\maketitle
The sign problem is a fundamental issue in Euclidean lattice field
theories at non-zero chemical potential, manifesting as complex weights
in the path integral \cite{deForcrand:2010ys,Gupta:2011ma,Aarts:2013bla}.
In this letter, we provide a technique that maps a large class of
Abelian lattice models with complex weights to dual models with real
weights. The dual form of these models can then be studied using familiar
analytical and computational methods. The models in this class possess
a generalized $\mathcal{PT}$ symmetry. In recent years, substantial
progress has been made in the study of models with this symmetry \cite{Bender:2005tb,Bender:2007nj,Meisinger:2012va}.
The methods developed here are applicable to models with a non-zero
chemical potential or a Minkowski-space electric field, which also
has a sign problem \cite{Shintani:2006xr,Alexandru:2008sj}. The utility
of lattice duality for the sign problem was shown some time ago \cite{Patel:1983qc,Patel:1983sc,DeGrand:1983fk},
and has recently been systematically studied \cite{Mercado:2011ua,Gattringer:2011gq,Mercado:2012yf,Mercado:2012ue,Gattringer:2012jt,Delgado:2012tm}
in an intermediate form, particularly in connection with the worm
algorithm \cite{Prokof'ev:2001zz}. The explicit duality relations
we derive here based on generalized $\mathcal{PT}$ symmetry represent
a solution to the sign problem for Abelian lattice models over a wide
range of parameter space. The dual forms generalize the well-known
chiral $Z(N)$ and Frenkel-Kontorova models and typically have a rich
phase structure with spatially-modulated phases \cite{Huse_PhysRevB.24.5180,ostlund_incommensurate_1981,yeomans_many_1981,yeomans_low-temperature_1982,bak_commensurate_1982}.
Such phases are also known to occur in $(1+1)$-dimensional fermionic
models \cite{Fischler:1978ms,Schon:2000he,Schnetz:2005ih,Basar:2008im,Basar:2008ki,Basar:2009fg},
and would also appear naturally in a quarkyonic phase \cite{Kojo:2009ha,Kojo:2011cn}.
However, spatially-modulated phases are not special to fermions at
finite density, as shown by a continuum model of $\left(1+1\right)$-dimensional
QCD with heavy particles where the statistics of the particles is
immaterial \cite{Ogilvie:2008zt,Ogilvie:2011mw,Meisinger:2012va}.
The appearance of spatially-modulated phases is natural in $\mathcal{PT}$-symmetric
models \cite{Ogilvie:2011mw,Meisinger:2012va}. 

In the models discussed here, the fundamental fields are elements
$z=\exp(i\theta)$ of $Z(N)$ or $U(1)$. The lattice actions are
complex, but invariant under the simultaneous application of the operators
$\mathcal{C}$ and $\mathcal{T},$ where $\mathcal{C}$ is a linear
charge conjugation operator that takes $\theta$ to $-\theta$, and
hence $z$ to $z^{*}$ ,and $\mathcal{T}$ is time reversal implemented
as complex conjugation. Thus these models have $\mathcal{CT}$ symmetry
as a generalized $\mathcal{PT}$ symmetry. In a lattice model, this
symmetry ensures that the eigenvalues of the transfer matrix are either
real or occur in complex conjugate pairs. The presence of complex
eigenvalues gives rise to a rich phase structure not possible with
Hermitian transfer matrices. Broadly speaking, there are three possibilities:
all eigenvalues are real (region I); the dominant eigenvalue of the
transfer matrix is real, but other eigenvalues form complex pairs
(region II); the dominant eigenvalues form a complex conjugate pair
(region III). It is known that models in region I are equivalent to
a Hermitian theory \cite{Mostafazadeh:2003gz}. In regions II and
III, the occurrence of complex conjugate pairs gives rise to spatially-modulated
behavior \cite{Ogilvie:2008zt,Ogilvie:2011mw,Meisinger:2012va}. Consider
the transfer matrix $T$ of a lattice model with $Z(N)$ or $U(1)$
variables such that $\mathcal{C}T\mathcal{C}=T^{*}$. The Fourier
transform operator $\mathcal{F}$ is a unitary matrix satisfying $\mathcal{F}^{T}=\mathcal{F}$
and $\mathcal{F}^{2}=\mathcal{F}^{+2}=\mathcal{C}$. $\mathcal{F}$
acts on $T$ to give a real transfer matrix $\tilde{T}=\mathcal{F}T\mathcal{F}^{+}$
satisfying $\tilde{T}^{*}=\tilde{T}$ \cite{Meisinger:2012va}. The
use of the Fourier transform in the construction of a real transfer
matrix is closely related to the use of lattice duality transformations
in finding real actions for models invariant under $\mathcal{CT}$. 

We begin with duality for $d=2$ $Z(N)$ models with a chemical potential
using the methods of \cite{Elitzur:1979uv} for the Villain, or heat
kernel, action. Defining the site-based spin variables as $\exp\left(2\pi im(x)/N\right)$,
with $m(x)$ an integer between $0$ and $N$, the partition function
is given by
\begin{equation}
Z[J,\mu\delta_{\nu,2}]=\sum_{m}\sum_{n_{\nu}}\exp\left[-\frac{J}{2}\sum_{x,\nu}\left(\frac{2\pi}{N}\partial_{\nu}m\left(x\right)-i\mu\delta_{\nu2}-2\pi n_{\nu}\left(x\right)\right)^{2}\right]
\end{equation}
where $\partial_{\nu}m(x)\equiv m(x+\hat{\nu})-m(x)$ and the sum
over link variables $n_{\nu}(x)\in Z$ ensures periodicity. Using
the properties of the Villain action, we can write

\begin{equation}
Z[J,\mu\delta_{\nu,2}]=\left(2\pi J\right)^{-dV/2}\sum_{m}\sum_{p_{\nu}}\exp\left[-\frac{1}{2J}\sum_{x,\nu}p_{\nu}^{2}\left(x\right)+i\sum_{x,\nu}p_{\nu}\left(x\right)\left(\frac{2\pi}{N}\partial_{\nu}m\left(x\right)-i\mu\delta_{\nu2}\right)\right]
\end{equation}
where $V$ is the number of sites on the lattice such that $dV$ is
the number of links. Summation over the $m\left(x\right)$'s give
a set of delta function constraints:

\begin{equation}
Z[J,\mu\delta_{\nu,2}]=\left(2\pi J\right)^{-dV/2}\sum_{p_{\nu}}\exp\left[-\frac{1}{2J}\sum_{x,\nu}p_{\nu}^{2}\left(x\right)+\sum_{x,\nu}p_{2}\left(x\right)\mu\right]\prod_{x}\delta_{\partial\cdot p,0\left(N\right)}
\end{equation}
where the notation in the Kronecker delta function indicates $\partial\cdot p=0$
modulo $N$. We introduce a dual bond variable $\tilde{p}_{\rho}\left(X\right)$
associated with the dual lattice via $p_{\nu}\left(x\right)=\epsilon_{\nu\rho}\tilde{p}_{\rho}\left(X\right)$
and note that the constraint on $p_{\nu}$ is solved by $\tilde{p}_{\rho}\left(X\right)=\partial_{\rho}\tilde{q}\left(X\right)+N\tilde{r}_{\nu}\left(X\right)$.
We have
\begin{equation}
Z[J,\mu\delta_{\nu,2}]=\left(2\pi J\right)^{-dV/2}\sum_{\tilde{q}}\sum_{\tilde{r}_{\nu}}\exp\left[-\frac{1}{2J}\sum_{x,\nu}\left(\partial_{\rho}\tilde{q}\left(X\right)+N\tilde{r}_{\nu}\left(X\right)\right)^{2}+\mu\sum_{x,\nu}\left(\partial_{1}\tilde{q}\left(X\right)+N\tilde{r}_{1}\left(X\right)\right)\right]
\end{equation}
which leads to
\begin{equation}
Z[J,\mu\delta_{\nu,2}]=\left(2\pi J\right)^{-dV/2}\exp\left[+\frac{V}{2}J\mu^{2}\right]Z[\frac{N^{2}}{4\pi^{2}J},-i\frac{2\pi J\mu}{N}\delta_{\nu,1}]
\end{equation}
 The generalized duality here is
\begin{eqnarray}
J & \rightarrow & \tilde{J}=\frac{N^{2}}{4\pi^{2}J}\\
\mu\delta_{\nu,2} & \rightarrow & \tilde{\mu}\delta_{\nu,1}=-i\frac{2\pi J\mu}{N}\delta_{\nu,1}.
\end{eqnarray}

The dual of the original model, which has a complex action, is a chiral
$Z(N)$ model with a real action; such models have been extensively
studied in two and three dimensions \cite{Huse_PhysRevB.24.5180,ostlund_incommensurate_1981,yeomans_many_1981,yeomans_low-temperature_1982}.
It is convenient to define a parameter $\Delta=J\mu$; the essential
characteristics can be understood by considering the range $0\le\Delta\le1$
\cite{ostlund_incommensurate_1981}. In the limit $\tilde{J}\rightarrow\infty$,
\emph{i.e.}, $J\rightarrow0$, configurations with $\partial_{\rho}\tilde{q}\left(X\right)=0$
are favored for $\Delta<1/2$; this leads to an extension of the ordered
phase of the dual model at $\Delta=0$ to non-zero $\Delta$. Beyond
$\mbox{\ensuremath{\Delta}=1/2}$, configurations with $\partial_{\rho}\tilde{q}\left(X\right)\ne0$
are favored in the same limit. In two dimensions, this corresponds
to phases with a nonzero value of the current coupled to $\mu.$ Similar
behavior will occur in a broad class of $Z(N)$ models, generalizable
to any dimension. In the specific case of a $d=2$ chiral $Z(N)$
model, an incommensurate spatially-modulated phase is found in the
$\tilde{J}-\Delta$ plane.

In the interesting case of the $Z(N)$ Villain Higgs model in $d=3$,
the partition function has the form 
\begin{eqnarray}
Z[J,K,\mu_{\nu},G_{\nu\rho}] & = & \sum_{m}\sum_{n_{\nu}}\sum_{p_{\nu}}\sum_{q_{\nu\rho}}\exp\left[-\frac{J}{2}\sum_{x,\nu}\left(\frac{2\pi}{N}\partial_{\nu}m\left(x\right)-\frac{2\pi}{N}p_{\nu}-i\mu_{\nu}-2\pi n_{\nu}\left(x\right)\right)^{2}\right]\nonumber \\
 &  & \times\exp\left[-\frac{K}{2}\sum_{x,\nu>\rho}\left(\frac{2\pi}{N}\left(\partial_{\nu}p_{\rho}-\partial_{\rho}p_{\nu}\right)-iG_{\nu\rho}-2\pi q_{\nu\rho}\right)^{2}\right]
\end{eqnarray}
where $\mu_{\nu}$ is a constant imaginary background vector gauge
field that generalizes the chemical potential, and $G_{\nu\rho}$
is a constant imaginary background field. This model is dual under
\begin{eqnarray}
J & \rightarrow & \tilde{J}=\frac{N^{2}}{4\pi^{2}K}\\
K & \rightarrow & \tilde{K}=\frac{N^{2}}{4\pi^{2}J}\\
\mu_{\nu} & \rightarrow & \tilde{\mu}_{\nu}=-i\frac{2\pi K}{N}\epsilon_{\nu\rho\sigma}G_{\rho\sigma}\\
G_{\nu\rho} & \rightarrow & \tilde{G}_{\nu\rho}=-i\frac{2\pi J}{N}\epsilon_{\nu\rho\sigma}\mu_{\sigma}
\end{eqnarray}
generalizing the well-known self-duality of the $d=3$ Abelian Higgs
system. The $d=3$ $Z(N)$ gauge field is dual to the $d=3$ chiral
$Z(N)$ spin model, which has been extensively studied \cite{yeomans_many_1981,yeomans_low-temperature_1982}.
In the strong-coupling limit where $K$ is small and thus $\tilde{J}$
large, the response of the system to an external real (Minkowski-space)
electric field reveals an infinite number of commensurate inhomogeneous
phases separating the disordered, confining phase of the gauge theory
from a phase with a constant induced field.

\begin{figure}
\includegraphics[width=3in]{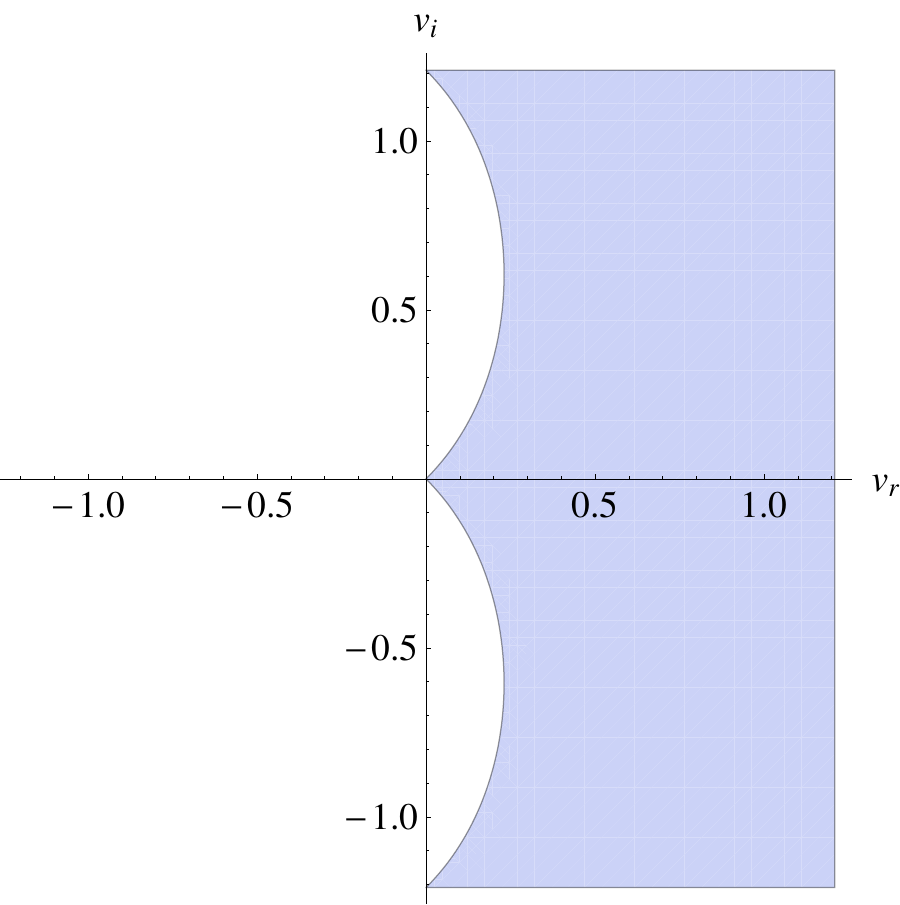}

\caption{\label{fig:Figure1}The Positive Dual Weight (PDW) region for a $Z(3)$
interaction as a function of $v_{r}$ and $v_{i}$.}
\end{figure}

The duality between $\mathcal{CT}$-symmetric interactions and chiral
interactions is not restricted to the Villain action, but holds more
generally. Consider a $Z(N)$ model on a $d$-dimensional lattice
with $N\ge3$ and local interactions on sites, links, plaquettes \emph{et
cetera}. A typical term in the lattice action is a function $V(z)$
where $z\in Z(N)$. It can be expanded as
\begin{equation}
V(z)=\sum_{j=0}^{N-1}v_{j}z^{j}.
\end{equation}
 We define $V_{j}=V(\omega^{j})$ where $\omega=\exp\left(2\pi i/N\right)$.
The $\mathcal{CT}$ operator takes $V(z)$ into $V^{*}(z^{*}).$ For
a $\mathcal{CT}$-symmetric interaction, this implies the $v_{j}$'s
are real and $V_{N-j}=V_{j}^{*}$ . A duality transform on $V$ is
implemented as a Fourier transform 
\begin{equation}
\exp\left(-\tilde{V}_{j}\right)=\sum_{k=0}^{N-1}\omega^{jk}\exp(-V_{k}).
\end{equation}
$\mathcal{CT}$ symmetry thus implies that the dual weights $\exp\left(-\tilde{V}_{j}\right)$
are real. If the dual weights are all positive, we can expand the
dual interaction as 
\begin{equation}
\tilde{V}(w)=\sum_{j=0}^{N-1}\tilde{v}_{j}w^{j}
\end{equation}
and we must have $\tilde{v}_{j}=\tilde{v}_{N-j}^{*}$, but the $\tilde{v}_{j}$'s
need not be real. This in general induces a chirality, a handedness
to the interactions, generalizing the chiral $Z(N)$ models to a larger
class of lattice models. There are large regions of parameter space
for which all the dual weights are positive, in which case we say
the model is in the positive dual weight class (PDW) class. Such models
may be simulated by standard computational methods such as the Metropolis
algorithm, and familiar theoretical tools such as mean field theory
may be applied. This represents a solution of the sign problem for
Abelian lattice models in the PDW class. Bochner's theorem states
that the strict positivity of $\exp\left(-\tilde{V}\right)$ is equivalent
to requiring that $\exp\left(-V\right)$ be positive-definite. This
in turn leads to the requirement $V_{0}<\left(V_{j}+V_{N-j}\right)/2$
for $1\le j\le N-1$, immediately excluding antiferromagnetic interactions
from the PDW class. Within the PDW region, the Perron-Frobenius theorem
applies to the dual representation of the transfer matrix, so there
is a single dominant eigenvalue. Thus the PDW region is disjoint from
region III. Figure \ref{fig:Figure1} shows the PDW region for a $Z(3)$
interaction in term of the variables $v_{r}=v_{1}+v_{2}$ and $v_{i}=v_{1}-v_{2}$.
The behavior shown is periodic in $v_{i}$ with a period of $2\pi/3\sqrt{3}\approx1.2092$.
In several models, the PDW region completely includes the region associated
with a non-zero chemical potential.

Many powerful techniques can be applied within the PDW region, including
low- and high-temperature expansions, variational methods such as
mean field theory, and renormalization group analysis. One particularly
powerful analytic method for spatially modulated phases combines mean
field theory in $d-1$ directions with the transfer matrix in the
spatially modulated direction \cite{mccullough_mean-field_1992}.
Correlation functions for local operators in the original complex
representation of the model can be constructed in the dual theory
in a well-known way \cite{PhysRevB.3.3918,Jose:1977gm}.

Further insight can be obtained from models based on $U(1)$. Here
we apply the duality techniques pioneered by Jose \emph{et al.} \cite{Jose:1977gm}.
The partition function of the two-dimensional XY model with an imaginary
chemical potential term has the form
\begin{equation}
Z[K,\mu\delta_{\nu,2}]=\int_{S^{1}}\left[d\theta\right]\sum_{n_{\nu}}\exp\left[-\frac{K}{2}\sum_{x,\nu}\left(\partial_{\nu}\theta\left(x\right)-i\mu\delta_{\nu2}-2\pi n_{\nu}\left(x\right)\right)^{2}\right].
\end{equation}
Again using the properties of the Villain action, we have 
\begin{equation}
Z[K,\mu\delta_{\nu,2}]=\int_{S^{1}}\left[d\theta\right]\prod_{x,\nu}\sum_{p_{\nu}\left(x\right)\in Z}\frac{1}{\sqrt{2\pi K}}e^{-p_{\nu}^{2}\left(x\right)/2K}e^{ip_{\nu}\left(x\right)\left(\nabla_{\nu}\theta\left(x\right)-i\delta_{\nu2}\mu\right)}.
\end{equation}
Integration over the $\theta$ variables leads to the constraint$ $$\nabla_{\nu}p_{\nu}\left(x\right)=0$.
This in turns allows us to write $p_{\rho}(x)=\epsilon_{\rho\nu}\nabla_{\nu}m(X)$
where $m\left(X\right)$ is an integer-valued field on the dual lattice
site $X$ which is displaced from $x$ by half a lattice spacing in
each direction. The partition function is now
\begin{equation}
Z=\sum_{\{m\left(X\right)\}\in Z}\frac{1}{\sqrt{2\pi K}}e^{-\sum_{X}\left[\sum_{\nu}\left(\nabla_{\nu}m\left(X\right)\right)^{2}/2K+\mu\nabla_{1}m\left(X\right)\right]}.
\end{equation}
The final step is to introduce a new field $\phi(x)\in R$ using a
periodic $\delta$-function, effectively performing a Poisson resummation:
\begin{equation}
Z=\int_{R}\left[d\phi\left(X\right)\right]e^{-\sum_{X}\left[\sum_{\nu}\left(\nabla_{\nu}\phi\left(X\right)\right)^{2}/2K+\mu\nabla_{1}\phi\left(X\right)\right]}\sum_{\{m\left(X\right)\}\in Z}e^{2\pi im\left(X\right)\phi\left(X\right)}.
\end{equation}
If we keep only the $m=1$ contributions, we have a lattice sine-Gordon
model
\begin{equation}
Z=\int_{R}\left[d\phi\left(X\right)\right]\exp\left[-\sum_{X,\mu}\frac{1}{2K}\left(\nabla_{\mu}\phi\left(X\right)\right)^{2}-\sum_{X}\mu\nabla_{1}\phi\left(X\right)+\sum_{X}2y\cos\left(2\pi\phi\left(X\right)\right)\right]
\end{equation}
with $y=1$. This will be recognized as a two-dimensional lattice
version of the Frenkel-Kontorova model, a sine-Gordon model with an
additional term proportional to $\mu$. For each fixed value of $X_{2}$,
the term $\sum_{X}\nabla_{1}\phi\left(X\right)$ counts the number
of kinks on that slice: The particles in the original representation
manifest as lattice kinks in the dual representation. This generalizes
to other lattice models based on $U(1)$, and can also be applied
to $Z(N)$ models realized by explicit breaking of $U(1)$ down to
$Z(N)$. From a continuum point of view, this model can be further
mapped to a massive Thirring model with $\mu$ coupling to the conserved
fermion current.

All of the Abelian lattice models in the $\mathcal{CT}$-symmetric
class studied here have real dual representations. These models typically
exhibit a rich phase structure in regions of parameter space where
the dual weights are positive. The properties of these models can
be studied in the dual representation with both computational and
analytical tools. Spatially-modulated phases can be detected in simulations
using appropriate two-point functions; analytical studies combined
with known results from condensed matter physics can provide valuable
guidance. Patel has recently suggested that an oscillatory signal
might appear in baryon number correlators in heavy ion collisions
at RHIC and the LHC \cite{Patel:2011dp,Patel:2012vn}. We believe
that the complex phase structure seen in Abelian systems is likely
to appear in non-Abelian systems. As an intermediate step, application
of duality to an effective Abelian model associated with the reduction
of $SU(N)$ to $U(1)^{N-1}$ \cite{Myers:2007vc,Unsal:2007vu,Unsal:2008ch,Ogilvie:2012is}
appears possible with the results developed here.
\begin{acknowledgments}
MCO thanks the U.S. Department of Energy for financial support, and
the Aspen Center for Physics for hospitality during the completion
of this manuscript.
\end{acknowledgments}
\bibliographystyle{unsrt}
\bibliography{\string"PT and Duality\string"}

\end{document}